\documentstyle[11pt,aaspp4]{article}

\def\parcsec{{\tt ''}\mskip -7.6mu.\,}   
\def\parcmin{{\tt '}\mskip -6.0mu.\,}   
\def\pdegree{^{\circ}\mskip -7.6mu.\,}  

\def\etal{{et~al.\ }}
\def\ie{{i.e.,}\ }

\def\refit{{\null}}

\def\simlt{\ {\raise-.5ex\hbox{$\buildrel<\over\sim$}}\ }
\def\simgt{\ {\raise-.5ex\hbox{$\buildrel>\over\sim$}}\ }
\def\kms{{km~s$^{-1}$}}
\def\subsun{_\odot}

\begin{document} 

\title{ Intracluster Planetary Nebulae in the 
Virgo Cluster 
\break I. Initial Results}

\author{John J. Feldmeier\altaffilmark{1} and Robin Ciardullo\altaffilmark{1}}
\affil{Department of Astronomy and Astrophysics, Penn State University,
525 Davey Lab, University Park, PA 16802}

\and

\author{George H. Jacoby}
\affil{Kitt Peak National Observatory, P.O. Box 26732, Tucson, AZ 85726}

\altaffiltext{1} {Visiting Astronomer, Kitt Peak National Optical Astronomy
Observatories, which is operated by the Association of Universities for 
Research in Astronomy, Inc., under cooperative agreement with the National 
Science Foundation.} 

\begin{abstract}
We report the initial results of a survey for intracluster planetary 
nebulae in the Virgo Cluster.  In two $16' \times  16'$ 
fields, we identify 69 and 16 intracluster planetary nebula candidates, 
respectively.  In a third $16' \times  16'$ field near the central 
elliptical galaxy M87, we detect 75 planetary nebula candidates, of which 
a substantial fraction are intracluster in nature.  By examining the
number of the planetaries detected in each field and the shape of
the planetary nebula luminosity function, we show that 1) the 
intracluster starlight of Virgo is distributed non-uniformly, and 
varies between subclumps A and B{}, 2) the Virgo Cluster core
extends $\sim 3$~Mpc in front of M87, and thus is elongated along 
the line-of-sight, and 3) a minimum of 22\% of Virgo's stellar 
luminosity resides between the galaxies in our fields, 
and that the true number may be considerably larger.  We also
use our planetary nebula data to argue that the intracluster
stars in Virgo are likely derived from a population that is of
moderate age and metallicity.
\end{abstract}


\section{Introduction}

The concept of intracluster starlight was first proposed by Zwicky (1951),
when he claimed to detect excess light between the galaxies of the Coma
cluster.  Follow-up photographic searches for intracluster luminosity in
Coma and other rich clusters (Welch \& Sastry 1971; Melnick, White, \& 
Hoessel 1977; Thuan \& Kormendy 1977) produced mixed results, and it was 
not until the advent of CCDs that more precise estimates of the amount of
intracluster starlight were made (cf.~Guldheus 1989; Uson, Boughn, \& Kuhn 
1991; V\'ichez-G\'omez, Pell\'o, \& Sanahuja 1994; Bernstein \etal 1995).
All these studies suffer from a fundamental limitation: the extremely
low surface brightness of the phenomenon.  Typically, the surface
brightness of intracluster light is less than 1\% that of the sky, and
measurements of this luminosity must contend with the problems
presented by scattered light from bright objects and the contribution of
discrete sources below the detection limit.  Consequently, obtaining
detailed information on the distribution, metallicity, and kinematics of
intracluster stars through these types of measurements is extremely
difficult, if not impossible.

An alternative method for probing intracluster starlight is through
the direct detection and measurement of the stars themselves.  Recent
observations have shown this to be possible.  In their radial velocity
survey of 19 planetary nebulae (PN) in the halo of the Virgo Cluster
galaxy NGC~4406 (M86), Arnaboldi \etal (1996) found 3 objects with
$v > 1300$~\kms; these planetaries are undoubtably intracluster in origin.
Similarly, Ferguson, Tanvir, \& von Hippel (1998) detected Virgo's
intracluster component via a statistical excess of red star counts in a
Hubble Space Telescope (HST) Virgo field over that in the Hubble Deep Field.
Finally, intracluster stars have been unambiguously identified from the
ground via planetary nebula surveys in Fornax (Theuns \& Warren 1997) and
Virgo (M\'endez \etal 1997; Ciardullo \etal 1998).

Motivated by these results, we have begun a large scale [O~III]
$\lambda 5007$ survey of intergalactic fields in Virgo, with the goal of
mapping out the distribution and luminosity function of intracluster
planetary nebulae (IPN){}.  Depending on the efficiency of tidal stripping,
Virgo's intracluster component is predicted to contain 
anywhere from 10\% to 70\% of
the cluster's total stellar mass (Richstone \& Malumuth 1983; Miller 1983).
A survey of several square degrees of Virgo's intergalactic space with
a four meter class telescope should therefore detect several 
thousand PN, and shed light on both the physics of tidal-stripping 
and on the initial conditions of cluster formation.  Here, we present 
the first results of our survey.

\section{Observations and Reductions}
On 1997 March 6-9 and 16, we imaged the Virgo Cluster through a 
44~\AA\ wide redshifted [O~III] $\lambda 5007$ filter (central wavelength,
$\lambda_c = 5027$~\AA) and a 267~\AA\ wide off-band filter ($\lambda_c =
5300$~\AA) with the prime focus camera of the Kitt Peak 4-m telescope and
the T2KB $2048 \times 2048$ Tektronix CCD{}.  In this initial survey, three
$16\arcmin \times 16\arcmin$ fields were chosen for study: one located at 
the isopleth center of Virgo subclump~A approximately $52\arcmin$ north and
west from M87, one $\sim 34\arcmin$ north and east of the 
giant elliptical NGC~4472 in subclump~B{}, and one $\sim 13\parcmin 8$ 
north of M87, also in subclump~A{} (For the definitions of the 
subclumps, see \cite{bts1}).  The conditions 
during the bulk of the observations were variable in both seeing and 
transparency.  However, at least one image of each field was taken 
on the photometric nights of 1997 Mar 8 and 9.  These photometric 
images served to calibrate the remaining frames, and
allowed us to put all our measurements on a standard system.

The exact coordinates of the field centers, as well as a summary of the
observations appear in Table~1.  Figure~1 displays the locations of our 
three fields, along with the locations of other detections of Virgo's
intracluster stars: M86 (via PN spectroscopy; Arnaboldi \etal 1996),
M87 (via PN imaging; Ciardullo \etal 1998), an HST Virgo field
(via excess red star counts; Ferguson, Tanvir, \& von Hippel 1998),
and a blank Virgo field observed with the William Hershel Telescope
(via PN imaging; M\'endez \etal 1997).

Our survey technique was as described in Jacoby \etal (1989) and Ciardullo,
Jacoby, \& Ford (1989b).  Planetary nebula candidates were identified by 
``blinking'' the sum of the on-band images against corresponding offband sums, 
and noting those point sources which were only visible in [O~III]{}.  This
procedure netted 76 planetary nebula candidates in Field~1, 16~PN in 
Field~2, and 75 in Field~3.  The magnitudes of the IPN were then measured
relative to bright field stars via the IRAF version of DAOPHOT (Stetson 1987),
and placed on a standard system by comparing large aperture measurements 
of field stars on the Mar~8 and 9 images with similar measurements of 
Stone (1977) standard stars.  Finally, monochromatic fluxes for the PN were
computed using the techniques outlined in Jacoby, Quigley, \& Africano (1987),
Jacoby \etal (1989) and Ciardullo, Jacoby, \& Ford (1989b).  These were
converted to $m_{5007}$ magnitudes using:
\begin{equation}
m_{5007} = -2.5 \log F_{5007} - 13.74
\end{equation}
where $F_{5007}$ is in units of ergs cm$^{-2}$ s$^{-1}$.

We note here that our $m_{5007}$ magnitudes carry an additional uncertainty 
which is unique to IPN photometry.  In order to compare the monochromatic flux
of an emission line object with the flux of a continuum source (\ie a 
standard star), the transmission of the filter at the wavelength of the
emission line must be known relative to its total integrated transmission
(cf.~Jacoby, Quigley, \& Africano 1987).  For PN observations within other
galaxies, this quantity is known (at least, in the mean) from the recessional
velocity and velocity dispersion of the target galaxy.  However, for our
intergalactic PN survey, we do not know {\it a priori\/} what the kinematic
properties of the target objects are, and hence we do not know the mean 
wavelength of their redshifted [O~III] $\lambda 5007$ emission lines.  In
deriving our monochromatic [O III] $\lambda 5007$ magnitudes, we have assumed
that the velocity dispersion of the intracluster PN follows that of the
Virgo Cluster as a whole (cf.~Binggeli, Sandage, \& Tammann 1985), but this
may not be true.  Although the systematic error introduced by this 
uncertainty is small ($\sim 5\%$), the effect may be important for 
individual objects.

\subsection{Obtaining a Statistical Sample}
Before we can compare the numbers and luminosity functions of IPN, 
we must first determine the photometric completeness limit in each field, and
define a statistical sample of objects.  Because our data were taken in 
mostly blank fields, our ability to detect IPN was not a strong function 
of position.  We therefore used the results of Jacoby \etal (1989) and Hui
\etal (1993) and equated our limiting magnitude for completeness with a
signal-to-noise of 9.  This is approximately the location where the PNLF 
(which should be exponentially increasing at the faint end) begins to turn 
down.  The relative depth of each field derived in this way also agrees with 
that expected from measurements of the seeing and mean transparency on the
individual images.  

In addition to excluding faint sources, we must make one other modification
to the IPN luminosity function.  Any galaxy that is within, 
or adjacent to our data frames, can potentially contaminate our 
IPN sample with PN still bound to their parent galaxies.  Because PN
are relatively rare objects, this source of contamination is unimportant
for the small, low luminosity, dwarf galaxies that are scattered
throughout the Virgo Cluster.  However, PN in the halos of bright galaxies
can be mistaken for intracluster objects, and must be taken into
account.

In Fields 1 and 2, there is only one contaminating object of any
consequence, NGC~4425, a relatively bright ($0.2 L^*$) lenticular
galaxy on the extreme western edge of Field~1.  From numerous 
observations (examples include \cite{paper5}; 
\cite{rbc1}), PN very closely follow the spatial distribution 
of galaxy starlight.  We therefore used surface photometry measurements 
to exclude seven of our PN candidates that fell 
within $2\parcmin 4$ of NGC 4425's nucleus.  From the $B$-band 
surface photometry of Bothun \& Gregg (1990), this distance
corresponds to approximately five effective radii or seven disk scale 
lengths.  Without these objects, we have 69 and 16 
IPN candidates in Fields~1 and 2, respectively, and a 
total of 44 and 9 objects in the photometrically complete samples.  

For Field 3, which has a total of 47 objects in its photometrically 
complete sample, the situation is more complex.  
Because the field is only $13\parcmin 8$ from the center of M87, 
some fraction of the objects in this region are probably bound 
to the galaxy.  The angular distribution
of the 75 planetary candidates does roughly follow the gradient defined
by the M87 surface brightness measurements of Caon, Capaccioli, \& 
Rampazzo (1990), and this does suggest an association with the galaxy.
However, these data, by themselves, do not exclude the possibility
that a large fraction of our PN are intracluster in origin.  Indeed,
since intracluster stars contribute luminosity just as galactic stars do,
it is difficult to distinguish intracluster light from galactic light
from surface photometry alone.  The situation is further complicated by the 
recent discovery of an extremely large, highly elongated, low surface 
brightness halo which surrounds M87 and extends $\sim 15\arcmin$ on the sky 
(Weil, Bland-Hawthorn, \& Malin 1997).  This structure is so large, that it is
probably not all bound to the galaxy.  The complexity of the region makes it 
impossible to determine the precise number of galactic PN contaminating the 
intracluster counts of Field~3 using surface photometry.  We will 
return to this issue in \S 3 below.

We note that there is one other possible source of contamination to our
intracluster PN sample.  Any object that emits a large amount of flux in our
redshifted $\lambda 5007$ filter, but is undetectable in the offband filter
will be mistaken for a planetary nebula.  Thus, distant objects, such as 
gas-rich starburst galaxies or quasars, could have enough flux in
a redshifted emission line to be detected in our survey.  In practice,
however, this is extremely unlikely.  The emission lines of importance
are [O~II] $\lambda 3727$ at $z \sim 0.35$ and Ly$\alpha$ at $z \sim 3.1$.
From surveys of high-redshift quasars (Schmidt, Schneider, \& Gunn 1995),
we should expect much less than one $z = 3.1$ bright quasar 
in our three fields combined.  Additionally, since our PN candidates 
have point-like point-spread-functions, those $z \sim 0.35$ galaxies 
with linear extents greater than $\sim 10$~kpc should have all been excluded 
on the basis of their angular size.  These and similar arguments 
made by other authors (Theuns \& Warren 1997; M\'endez \etal 1997), 
imply that contamination from background sources is probably not 
significant in our survey.  

The final luminosity functions for our 3 fields are plotted in Figure~2.  For 
comparison, a sample of suspected intracluster PN's from M87's inner halo 
(Ciardullo \etal 1998) are also plotted.   
 
\section{The Distribution of Intracluster Stars}

The most obvious feature displayed in Figure~2 is the dramatically different
numbers of IPN in Fields 1 \& 2.  Although the survey depths of the three
regions differ (due to differences in sky transparency and seeing), it is
clear that the density of intracluster objects in Field~1 at 
the center of subclump~A{} is much larger than that in Field~2.  After 
accounting for the different depths, the PN density in Field 3, 
near M87, is smaller by a factor of at least two, and the number 
of PN in Field~2, which is near the edge of the 6 degree Virgo 
Cluster core in subclump~B, is down by a factor of $\sim 4$. This 
behavior is quite intriguing.

The fact that subclump~B has fewer PN than subclump~A can probably
be attributed to cluster environment.  It is well known that subclump~B
has fewer early-type galaxies than subclump~A (\cite{bts1}).  If ellipticals
and intracluster stars have a related formation mechanism (\ie galaxy
interactions), then a direct correlation between galaxy type and stellar
density in the intergalactic environment might be expected.

What is harder to understand is the high IPN density in Field~1.  Although
galaxy isopleths place the Virgo Cluster center in Field~1 (\cite{bts1}),
X-ray data clearly demonstrate that the true center of the cluster is at
M87 (B\"ohringer \etal 1994).  The reason for the offset is a subcluster of
galaxies associated with M86{}.  Kinematic data (Binggeli, Popescu, \& 
Tammann 1993) and ROSAT X-ray measurements (B\"ohringer \etal 1994)
both show that the center of Virgo subclump~A is contaminated by a separate
group of galaxies which is falling in from the far side of the cluster.  This
interpretation is confirmed via direct distance measurements using the
planetary nebula luminosity function (Jacoby, Ciardullo, \& Ford 1990)
and surface brightness fluctuation method (Ciardullo, Jacoby, \& Tonry 1993;
Tonry \etal 1997): both place M86 $\sim 0.3$~mag behind M87.  At this
distance, the PN associated with M86 and its surroundings should be beyond the
completeness limit of our survey, and should not be contributing to our 
PN counts.  The observed IPN density of Field~1 should therefore be 
smaller than that measured near M87, not greater.

A second feature of Figure~2 is the slow fall-off of the bright-end of
Field~3's planetary nebula luminosity function (PNLF).  Observations in 
$\sim 30$ elliptical, spiral, and irregular galaxies have demonstrated that a 
system of stars at a common distance will have a PNLF of the form:
\begin{equation}
N(M) \propto e^{0.307M} \, [1 - e^{3(M^{*}-M)}]
\end{equation}
where $M^* \approx -4.5$ (Jacoby \etal 1992).
{\it In no isolated galaxy has the PNLF ever deviated from this form.}
However, in a cluster environment such as Virgo, the finite depth of the
cluster can distort the PNLF, as PN at many different distances  
contribute to the observed luminosity function.  
Due to the lack of survey depth in Field~1, and small number of PN in
Field~2, we cannot determine whether the PNLFs of these fields differ 
from the empirical function.  However the PNLF of Field~3 is clearly
distorted, in the manner identical to that found by Ciardullo \etal (1998)
in their survey of the envelope of M87.

Figure~3 illustrates this effect in more detail.  In the figure, we first 
compare the observed PNLF of Field~3 to the most likely empirical curve 
(solid line) as found via the method of maximum likelihood (Ciardullo \etal 
1989a).  The fit is extremely poor, and a Kolmogorov-Smirnov test rejects the 
empirical law at the 93\% confidence level.  The reason for the poor fit
is simple:  the presence of bright, ``overluminous'' objects forces the 
maximum-likelihood technique to find solutions which severely
overpredict the total number of bright PN in the field.

There is only one plausible hypothesis for the distorted PNLF and the
overabundance of bright PN --- the presence of intracluster objects.  An 
instrumental problem is ruled out, since the bright-end distortion has also 
been seen in data taken of M87's envelope two years earlier with a different
filter and under different observing conditions (Ciardullo \etal 1998).
Similarly, extreme changes in metallicity cannot explain the discrepancy.
Unless the bulk of the intracluster stars are super metal-rich ([O/H] $\simgt
0.5$), abundance shifts can only decrease the luminosity of the [O~III] 
$\lambda 5007$ line, not increase it (Dopita, Jacoby, \& Vassiliadis 1992; 
Ciardullo \& Jacoby 1992).  Finally, neither population age nor the existence 
of dust is a likely scenario: the former requires an unreasonably young 
($< 0.5$~Gyr) age for M87's stellar envelope (Dopita, Jacoby, \& Vassiliadis 
1992; M\'endez \etal 1993; Han, Podsiadlowski, \& Eggleton 1994), and the 
latter implies a significant gradient in foreground extinction between 2 and 7
effective radii from M87's center.

Thus, one is left with the hypothesis, first presented by Jacoby (1996)
and Ciardullo \etal (1998), that intracluster PN are responsible 
for the distorted PNLF{}.
From our own study, and that of others (Arnaboldi \etal 1995; M\'endez 
\etal 1997; Ferguson, Tanvir, \& von Hippel 1998), it is clear that a 
significant fraction of intracluster stars exist in Virgo, and some of these 
stars should be positioned in front of M87 along the line-of-sight.  
The existence of foreground PN naturally explains the existence of
``overluminous'' PN, and is supported by the fact that the brightest PN
in Field~3 are of comparable magnitude to the brightest objects found
in the Ciardullo \etal (1998) survey of M87.

The distorted PNLF gives us a way to estimate a lower limit to the 
number of IPN in Field~3.  To do this, we plot a dashed line 
in Figure~3, which represents the expected luminosity function 
of M87 planetaries using the observed PNLF distance modulus of the 
inner part of the galaxy ($m-M = 30.87$; Ciardullo \etal 
1998).  To normalize this curve, we assume that {\it all\/} the PN at 
$m_{5007} = 26.9$ belong to the galaxy.  As is illustrated, there is an 
excess of bright objects compared to that expected from M87 alone: these are 
objects at the bright end of the intracluster planetary nebula population.
If we statistically subtract the M87 luminosity function from the Field~3
data, we arrive at the conclusion that at least $10 \pm 2$ planetaries 
with $m_{5007}$ brighter than 26.5 are intracluster in nature, where
the uncertainty is solely due to the uncertainty in the normalization 
value.  

Note that this estimate is a bare minimum for the number of intracluster
PN{}.  For all reasonable planetary nebula luminosity functions, there
are many more faint PN than bright PN{}.  Thus, our assumption that all
$m_{5007} = 26.9$~PN are galactic is clearly wrong.  However, without
some model for the distribution of intracluster stars, the shape of the
intracluster PNLF cannot be determined.  This makes it impossible to
photometrically distinguish faint intracluster PN from PN that are bound
to M87.  For the moment, we therefore conservatively claim that at least $10 
\pm 2$ intracluster planetaries are present in Field~3.  In the future, 
it should be possible to refine this estimate with improved 
observations of the precise shape of the PNLF, and with the use of 
dynamical information obtained from PN radial velocity measurements.

A final feature of Figure~2, and perhaps the most remarkable, deals with
distance.  Even a cursory inspection of Figure~2 shows that the IPN of Field~1
are significantly brighter than those of Field~2.   If we fit the two
distributions to the PNLF of equation (2) via the technique of maximum 
likelihood (Ciardullo \etal 1989a), then the most likely distance to the PN
of Field~1 is $11.8 \pm 0.7$~Mpc, while that for Field~2 is $14.7 \pm 1.5$~Mpc.
The fact that Field~2 is more distant is not surprising, since both Yasuda, 
Fukugita, \& Okamura (1997) and Federspiel, Tammann, \& Sandage (1997)
place the galaxies of subclump~B $\sim 0.4$~mag behind those of subclump A{}.
What is surprising is the relatively small distance to the IPN of Field~1.
Most modern distance determinations, including the analysis of Cepheids
in spirals by van den Bergh (1996) and the measurement of the planetary
nebula luminosity function and surface brightness fluctuations in ellipticals
(Jacoby, Ciardullo, \& Ford 1990; Tonry \etal 1997), place the core of the
Virgo Cluster at a distance of between 14 and 17~Mpc.  No modern measurement 
to Virgo gives a distance smaller than $\sim 14$~Mpc, and 11.8~Mpc is certainly
not a reasonable value for the distance to the cluster.

The reason for the $\sim 3$~Mpc discrepancy is that the PNLF law has an
extremely sharp cutoff at the bright end of the luminosity function.  As a
result, our PN detections are severely biased towards objects on the front
edge of the cluster, and our distance estimates carry the same bias.
Our derived value of 11.8~Mpc therefore represents the distance to the front 
edge of the Virgo Cluster, not the distance to the galaxies in the Virgo 
Cluster core.

Another way of looking at the data is to consider the brightest IPN in 
Fields~1 and 3.  If we assume that these bright objects are indeed part 
of the Virgo Cluster and have an absolute magnitude near $M^*$, then
their apparent magnitudes yield an immediate upper limit to the front edge
of each field.  The result is that the brightest IPN of Fields~1 and 3 have
distances of no more than $11.7$~Mpc, \ie they are $\sim 3$~Mpc in front of
the cluster core, as defined by the original PNLF measurements of Jacoby,
Ciardullo, \& Ford (1990).  The same conclusion was reached by Ciardullo
\etal (1998) using the sample of PN around M87's inner halo.  This distance
is significantly larger than the size of the cluster projected on the sky:
at a distance of $\sim 15$~Mpc, the classical 6~degree core of Virgo 
translates to a linear extent of only $\sim 1.5$~Mpc.  Although it is 
unclear how our measurement of cluster depth quantitatively compares to 
the classical angular estimate of the core 
(Shapley \& Ames 1926; see the discussion in de Vaucouleurs 
\& de Vaucouleurs 1973), our data does suggest that
Virgo is elongated along our line-of-sight, perhaps by as much as a factor
of two.

It is worth repeating here that the planetary nebula luminosity function is
extremely insensitive to the details of its parent population, and those
dependences that do exist cannot explain the bright apparent magnitudes
seen in the IPN population.  The problem is more fully discussed in
Ciardullo \etal (1998), but, in summary, there is no reasonable explanation
for the existence of these bright [O~III] $\lambda 5007$ sources other than
that their location is in the foreground of the Virgo Cluster.  We note that
many authors have suggested that the Virgo Cluster has substantial depth
(for example, see the galaxy measurements of Pierce \& Tully 1988; 
Tonry, Ajhar, \& Luppino 1990; Yasuda, Fukugita, \& Okamura 1997), 
but this direct measurement is still quite surprising.

\section{The Stellar Population and Total Number of Intracluster Stars}

Renzini \& Buzzoni (1986) have shown that bolometric-luminosity specific 
stellar evolutionary flux of non-star-forming stellar populations should 
be $\sim 2 \times 10^{-11}$~stars-yr$^{-1}$-$L_{\odot}^{-1}$, nearly 
independent of population age or initial mass function.  If the 
lifetime of the planetary nebula stage is $\sim 25,000$~yr, then 
every stellar system should have $\alpha \sim 50 \times 
10^{-8}$~PN-$L_{\odot}^{-1}$.  Observations in 
elliptical galaxies and spiral bulges have shown that no galaxy has a value 
of $\alpha $ greater than this number, but $\alpha $ can be a 
up to a factor of five smaller (\cite{rbciau}). 
Nevertheless, the direct relation between number of planetary 
nebulae and parent system luminosity does provide us with a tool with 
which to estimate the number of stars in Virgo's 
intergalactic environment.

In order for us to estimate the density of intracluster stars in Virgo, we
must first fit the observed PNLF with a model which represents the distribution
of planetary nebulae along the line-of-sight.  Such a model is a necessary
step in the analysis: as seen above, Virgo's intracluster component extends
$\sim 0.5$~mag in front of its core, and thus our sample of PN is severely 
biased towards objects on the near side of the cluster.  To investigate the 
effect of this bias, we considered two extreme models for the distribution
of Virgo's intracluster stars: a single component model, in which all the PN
are at a common distance, and a radially symmetric model, in which the 
intracluster stars are distributed isotropically throughout a sphere of radius 
3~Mpc.  The symmetric model was then adjusted to deliver a ``best fit'' to 
the observed PNLF at the assumed cluster distance of 15~Mpc (Jacoby, Ciardullo,
\& Ford 1990).  For the single component models, we adopted 
the distances derived in \S 3.  The most likely value for the 
underlying population's stellar luminosity was then calculated 
using the method of maximum likelihood (Ciardullo \etal 1989a) 
and an assumed value of $\alpha_{2.5} = 20 \times 10^{-9}$, which 
is an average value for elliptical galaxies ($\alpha_{2.5}$ is the 
number of PN within 2.5~mag of $M^*$ per unit bolometric 
luminosity).  Because of our limited knowledge of the
luminosity function of Field 3, we used single component models only for
this field, assuming that field contains 10 IPN within our completeness 
limit.  In addition, because of the very large uncertainties in 
the above models, we also computed a single component model 
using $\alpha_{2.5} = 50 \times 10^{-9}$~PN-$L_{\odot}^{-1}$; this last 
model represents the minimum amount of intracluster starlight necessary to
be consistent with our data.  The results of our models are summarized 
in columns 1-4 in Table~2.

The total amount of intracluster starlight found is quite large, at least
$6.8 \times 10^{9} L\subsun$ in the 768 square arcminutes we surveyed, and
probably much more.  As expected, the density of intracluster stars varies
significantly between the fields.  For Fields~1 and 2, the choice of cluster
model changes the amount of derived intracluster light dramatically.  In the 
single component model, most of the PN can be on the near side of the cluster,
where we can see objects relatively far down the luminosity function.
In this scenario, the size of the IPN's parent population is relatively small,
as is the amount of intracluster light.  In contrast, the symmetric model
places a large number of stars on the back side of the cluster, where they 
contribute light, but do not populate the bright end of the PN luminosity 
function.  The intracluster luminosity in this picture is correspondingly 
larger.

How likely is it that the intracluster stars are distributed symmetrically
in the cluster?  There is strong evidence to suggest that neither the galaxies 
nor the intracluster stars are in virial equilibrium.  Based on the 
distribution and kinematics of galaxies, Binggeli, Tammann, \& Sandage (1987)
and Bingelli, Popescu, \& Tammann (1993) concluded that the core of Virgo 
exhibits a significant amount of substructure.  Similarly, Ciardullo \etal
(1998) showed that the PNLF of intracluster stars near M87 is incompatible 
with that of a relaxed system.  Although these analyses do not formally
exclude all symmetric distributions, they do suggest that a symmetric
distribution is unlikely.  

Although our limited amount of photometric data do not allow us to address 
the question of Virgo's structure directly, we can gain some insight into 
the state of the intracluster stars by speculating about their likely origin.
To do this, we first focus on the metallicity of the observed IPN{}.  Many of
the planetary nebulae detected in our survey are extremely bright: some are
more than 0.6~mag brighter than $m^*$ at the center of the cluster.  In \S 3,
we interpreted this brightness in terms of distance, and were thus able to 
place the IPN $\sim 3$~Mpc in front of M87.  This argument, however, assumes 
that $M^*$, the absolute magnitude of the PNLF cutoff, is well known.  
For most stellar populations, this is a good assumption, as evidenced by the
agreement between PNLF distances and distances determined from other
methods (cf.~Jacoby \etal 1992; Ciardullo, Jacoby, \& Tonry 1993; Feldmeier,
Ciardullo, \& Jacoby 1997).  However, in extremely metal poor systems, the 
decreased number of oxygen atoms present in a PN's nebula does have 
an affect.  Specifically, PN from a population whose metallicity is 
one-tenth solar are expected to have a value of $M^*$ that is fainter 
than the nominal value by more than 0.25~mag (Ciardullo \& Jacoby 1992; 
Dopita, Jacoby, \& Vassiliadis 1992; Richer 1994).  This increase in 
$M^*$ translates directly into an error in distance.  
For example, if the intracluster environment of Virgo were filled with
stars with one-tenth solar abundance, our derived distance to the 
front of the cluster would be overestimated by $\sim 11\%$, and our 
value for the distance between the front of the cluster and the 
Virgo core galaxies would be underestimated by almost 50\%.
Note, however, that we already measure a Virgo Cluster depth that is 
$\sim 2$ times that of its projected size; lowering the metallicity of 
the stars would only increase this ratio.  It therefore seems likely that 
the PN detected in our survey are of moderate metallicity.

Similarly, the mere fact that we do see a large number of IPN suggests that
most of the intergalactic stars are not extremely old.  In their [O~III]
$\lambda 5007$ survey of Galactic globular clusters, Jacoby \etal (1997)
found a factor of $\sim 4$ fewer PN than expected from stellar evolution
theory.  Jacoby \etal attribute this small number to the extreme age of
the stars.  Observations in the Galaxy suggest that stars with turnoff
masses of $\sim 0.8 M\subsun$ produce post asymptotic giant branch stars
with extremely small cores, $M_c < 0.55 M\subsun$ (Weidemann \& Koester 1983).
Objects such as these evolve to the planetary nebula phase very slowly:  
so slowly, in fact, that their nebulae can diffuse into space before the 
stars become hot enough to produce a significant number of ionizing photons.  
If this scenario is correct, then the fact that we see large numbers 
of IPN indicates that the intracluster stars of Virgo cannot be as old 
as the globular cluster stars of the Galaxy.

If the intracluster stars of Virgo are, indeed, of moderate age and 
metallicity, then they must have been stripped out of their 
parent galaxies at a relatively recent epoch.  One likely way of 
doing this is through ``galaxy harassment'' whereby high-speed 
encounters between galaxies rip off long tails of matter
which lead and follow the galaxy (Moore \etal 1996).  Although this
tidal debris will eventually dissolve into the intracluster environment,
it is possible that the increased energy of these stars may cause them
to linger in the outer parts of the cluster for a long time.
It is therefore possible that the intergalactic environment of Virgo is 
not a homogeneous region, but is instead clumpy, and filled with 
filaments.  It may be that the bright PN present in Field~1 belong 
to one such structure that happens to be in the front side of the 
cluster.  If this is correct, the symmetric model for Field~1 should
be a gross overestimate of the true amount of intracluster starlight.     

If we assume that the intracluster stars come from an old stellar
population, similar to that found in M87 with a 
$B$$-$$V$ color of 1.0 (de Vaucouleurs \etal 1991), a bolometric 
correction of $-0.85$ (Jacoby, Ciardullo, \& Ford 1990), and 
a mean distance of 15~Mpc, then the luminosities implied by our PN 
observations can be expressed in terms of $B$-band surface brightnesses.
Although the values are position and model dependent (cf.~Table~2, 
column~5), the result is quite interesting.  Our PN observations imply
that the surface brightness in the Virgo Cluster core varies between
$B \sim 26$ and $B \sim 30$~mag per sq.~arcsec.  These values are in 
excellent agreement with other PN-derived surface brightness values 
in Virgo (M\'endez \etal 1997) and Fornax (Theuns \& Warren 1997).  They are, 
however, significantly larger than the value of $B \sim 31.2$ implied from
red star counts on {\sl HST\/} frames (Ferguson, Tanvir, \& von Hippel 1998). 
One possible explanation for this discrepancy lies in the locations of the 
fields: the {\sl HST\/} field is $50\arcmin$ east of M87, whereas Fields~1 and 
2 are in denser regions of the cluster.  If intracluster stars are concentrated
towards the cores of the subclumps of Virgo, then the low number of red stars
observed by {\sl HST\/} may be attributable to the field's location.  Under
this assumption, and using the same stellar population model used by
Ferguson, Tanvir \& von Hippel (1998), we would expect only $\sim 5$
IPN brighter than $m_{5007} = 27.0 $ in a $16\arcmin \times 16\arcmin$ field 
centered on the {\sl HST\/} position.

On the other hand, our fields may, indeed, be typical locations in the
Virgo Cluster.  As described above, the high galaxy density in
Field~1 is due, in part, to objects on the far side of the 
cluster which do not contribute to the observed sample of planetaries.  
Similarly, the shape of the PNLF of Field~3 suggests that much of the 
intracluster luminosity in that region is not associated 
with the physical core of Virgo, but is only present in the
field through projection.  

If our survey regions are typical of Virgo in general, then we can use
our observations to estimate the total fraction of Virgo's starlight 
which is between the galaxies.  By fitting the surface distribution of Virgo
galaxies in subclump~A to a King model with core radius $1\pdegree 7$, 
Binggeli, Tammann, \& Sandage (1987) derived a central luminosity density
of the cluster of $1 \times 10^{11} L\subsun$ per square degree.  If we 
scale this galactic luminosity density, which we denote as 
L$_{\it galaxies}$ (cf.~Table~2, column~6), to the sizes and locations of
Fields~1 and 3, we can directly determine the importance of intracluster
starlight.  Due to its irregularity, Binggeli, Tammann, \& Sandage (1987) did 
not fit an analytical model to subclump~B{}.  However, the luminosity
density in Field~2 is certainly not greater than that for Field~1.  We 
therefore use Field~1's value to set a lower limit on the fraction of 
intracluster starlight in Field~2.  The results are given in column 7 of 
Table~2.

Depending on the model and the field, the fraction of intracluster light
varies from 12\% to 88\% of the cluster's total luminosity.  To set
a lower limit to the relative importance of intracluster starlight, we 
average the results from our three fields, using the smallest 
fraction found for each field.  We find an average fraction of 22\%.  
Similarly, to find the upper limit to the intracluster fraction, we
average the largest fractions determined.  In this case, we find an 
average fraction of 61\%.  This range is in rough agreement 
with the results derived from other 
PN observations in Virgo (M\'endez \etal 1997) and Fornax 
(Theuns \& Warren 1997).  They also agree with the direct measurement 
of intracluster light in Coma (Bernstein \etal 1995).

We stress that these results are very uncertain, due to our lack of 
knowledge about the true distribution of intracluster stars, 
and the possible variation of $\alpha_{2.5}$.  In addition, it is 
also probable that we have missed a small fraction of the IPN present 
in our fields due to the finite width of our interference filter.   
If the IPN velocity dispersion follows that of the galaxies, then 
$\sim 8\%$ of the IPN will be Doppler shifted out of the bandpass of 
our 44 \AA\ full-width half-maximum filter.  However, since the
true kinematics of the IPN are unknown, this fraction cannot be determined
precisely.  Nevertheless, regardless of the particular model, a large
fraction of intracluster stars are present in the Virgo Cluster.

The large amount of intracluster starlight found by this and 
other studies places new constraints on models of the formation 
and evolution of galaxy clusters.  If, as we have suggested, the  
intracluster stars are of moderate age and metallicity, they must not
have been formed in the intracluster environment, but instead 
have been removed from their parent galaxies during encounters.  
Therefore, the number and distribution of intracluster stars could
be a powerful tool for discovering the history of individual galaxy clusters.  
Furthermore, since intracluster stars are already free of the 
potential wells of galaxies, they may contribute a significant fraction of 
metals and dust to the intracluster medium.

The large amount of intracluster stars is also an unrecognized source 
of baryonic matter that must be taken into account in studies of galaxy 
clusters.  Though not enough to account for all of the dark matter,
intracluster stars do increase the fraction of matter that is in baryonic
form.  This has potentially serious consequences for 
cosmological models.  From their calculation of the baryon fraction 
in the Coma Cluster, White \etal (1993) found that their derived value
was too large for the universe to simultaneously have $\Omega_0 = 1$
and also be in accord with calculations of cosmic nucleosynthesis.
This result, sometimes called the ``baryon catastrophe,'' has been
confirmed in several other galaxy clusters (White \& Fabian 1995;
Ettori, Fabian, \& White 1997).  Although the exact amount of intracluster
starlight is currently very uncertain, the presence of a large number of
intracluster stars can only increase the baryon discrepancy already found.
More observations will be necessary to establish how much intracluster
starlight adds to the observed baryon fraction of galaxy clusters.    

\section{Conclusion}

We report the results of a search in three fields in the Virgo Cluster 
for intracluster planetary nebulae, and have detected a total of 
95 intracluster candidates.  From analysis of the numbers of the 
planetaries, we find that the amount of intracluster light in Virgo 
is large (at least 22\% of the cluster's total luminosity), distributed 
non-uniformly, and varies between subclump~A and B{}.  By using the 
planetary nebulae luminosity function, we derive an upper
limit of $\sim 12$~Mpc for the distance to the front edge of the Virgo
Cluster and use this to show that the cluster must be elongated 
along our line of sight.  We also use the properties of planetary 
nebulae to suggest that the intracluster stars of Virgo have 
moderate age and metallicity.  The large fraction of intracluster stars 
found has potentially serious consequences for models of cluster 
formation and evolution, and for cosmological models.  Finally, 
we note that this survey included less than 0.2\% of the traditional 
6 degree core of the Virgo Cluster.  Many more intracluster planetary 
nebulae wait to be discovered.

We thank Allen Shafter for some additional off-band observations 
and Ed Carder at NOAO, for his measurements of our on-band filter 
so that we could begin our observations on time.  We would also
like to thank the referee, R. Corradi, for several suggestions 
that improved the quality of this paper.  Figure~1 was
extracted from the Digitized Sky Survey, which was produced
at the Space Telescope Science Institute under U.S. Government grant 
NAGW-2166.  This work was supported in part by NASA grant NAGW-3159
and NSF grant AST95-29270.

\pagebreak

\clearpage

\begin{figure}
\caption{A $2\pdegree 5$ by $5\pdegree 7$ region of the Virgo 
Cluster, drawn from the Digitized Sky Survey.  North is up, and 
east is to the left.  Our three
fields are indicated by the large squares of side $16\arcmin$.  Other
detections of intracluster stars are also displayed including the 
intracluster planetary nebulae in front of M86
(Arnaboldi \etal 1996) and M87 (Ciardullo \etal 1998), intracluster
red giant and asymptotic giant branch stars detected by the Hubble Space 
Telescope (Ferguson, Tanvir, \& von Hippel 1998), and the intracluster
planetary nebulae detected in the $4\arcmin$ radius circular field of 
M\'endez \etal (1997).  The position of NGC 4472 is shown
for reference.}  
\end{figure} 

\begin{figure}
\plotfiddle{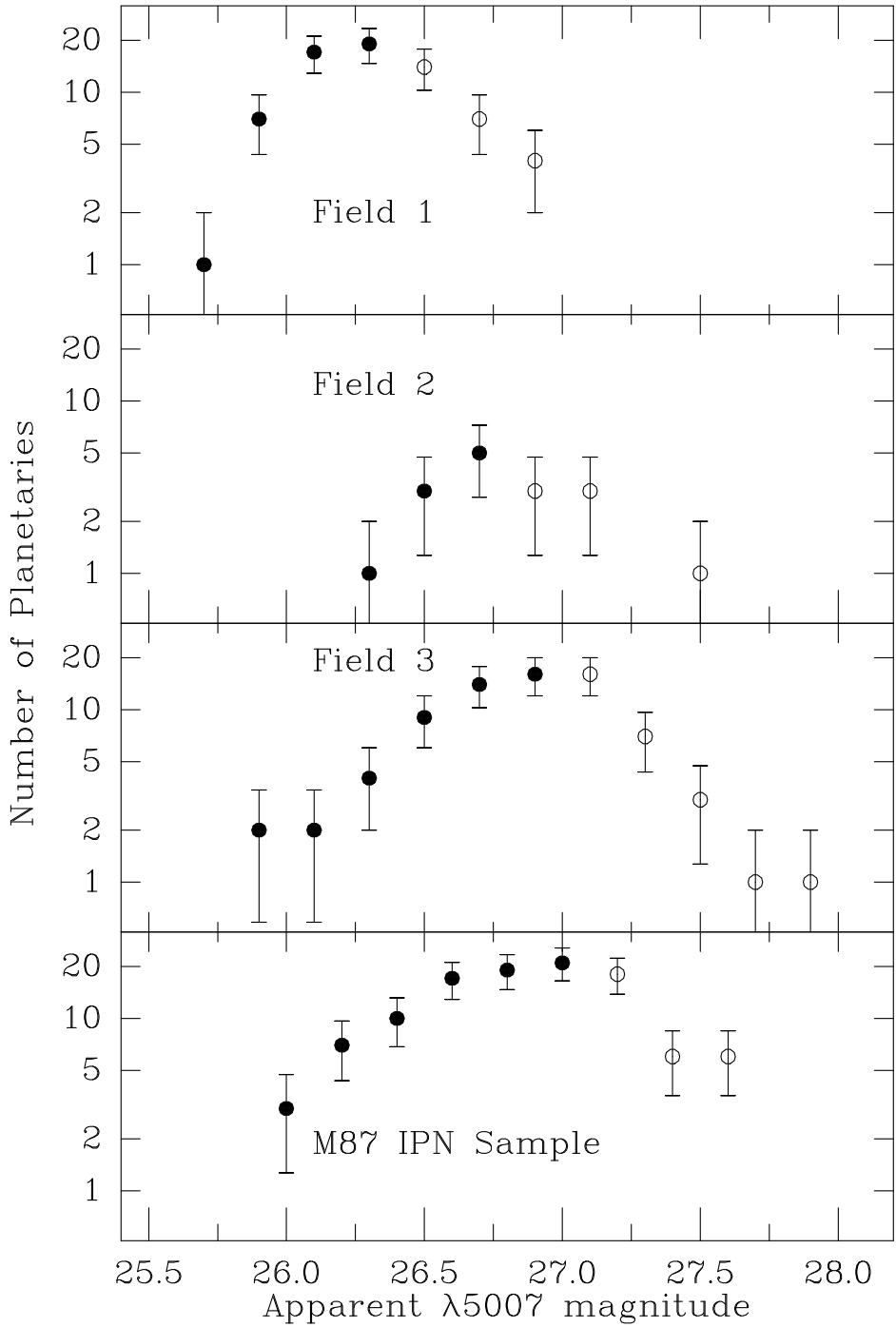}{500pt}{0}{85}{85}{-250}{-90}
\caption{The planetary nebula luminosity functions for the three intracluster
fields, plus a sample of suspected intracluster planetaries from M87
(Ciardullo \etal 1998), binned into 0.2~mag intervals.  The solid 
circles represent objects in our statistical IPN samples; the open
circles indicate objects fainter than the completeness limit.  Note 
that the numbers of intracluster planetaries vary from field to field,
and that Field~2 appears to have a fainter cutoff than Fields~1 and 3.}
\end{figure}

\begin{figure}
\plotfiddle{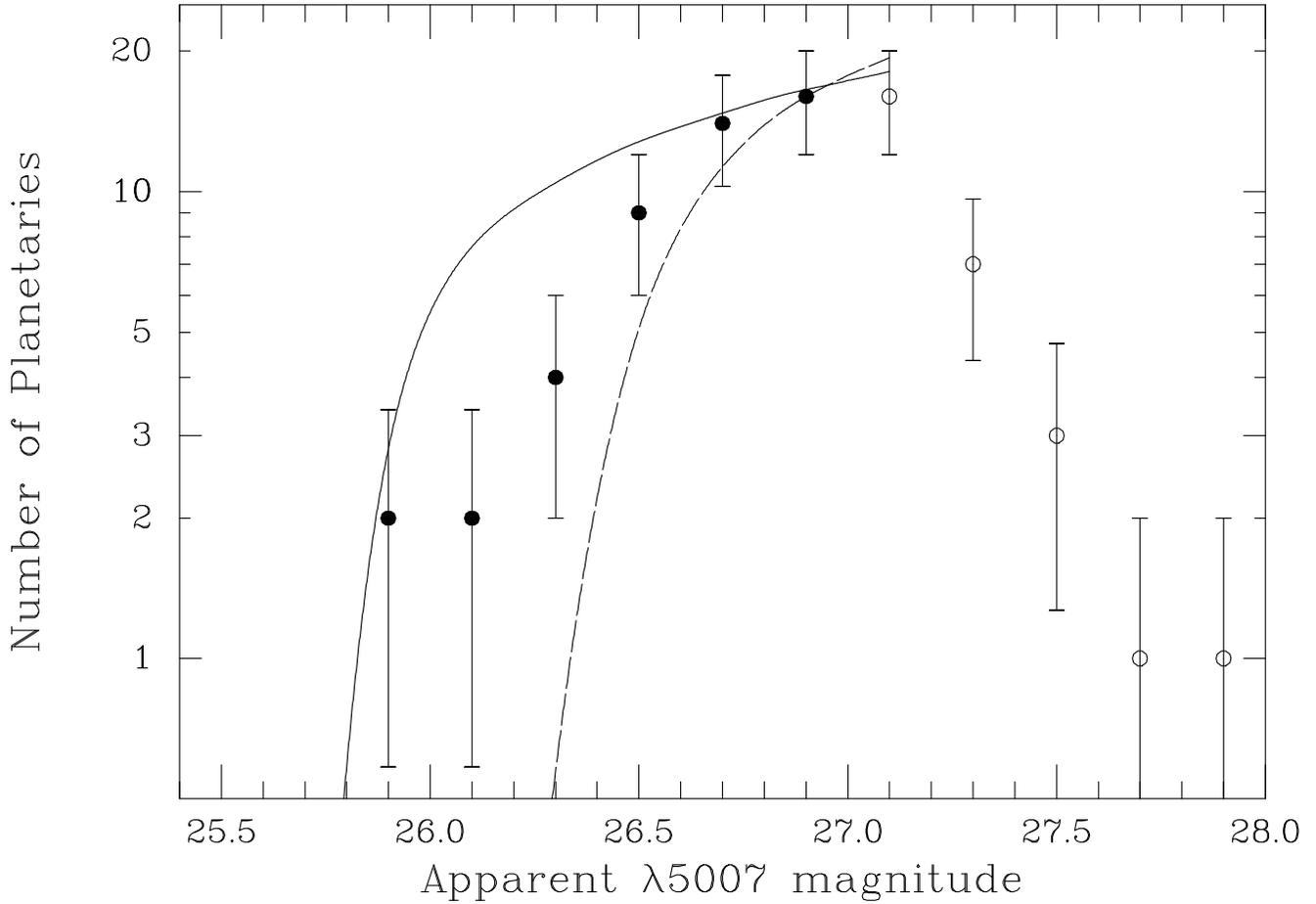}{250pt}{-90}{85}{85}{-300}{450}
\caption{The planetary nebula luminosity functions for Field~3, binned into
0.2~mag intervals.  The solid circles represent objects in the 
statistical sample; the open circles indicate objects fainter than the
completeness limit.  The solid line shows the empirical planetary nebulae
luminosity function shifted to the most likely distance modulus --- it
is excluded at the 93\% confidence level.  The dashed line is the 
planetary nebulae luminosity function placed at the best-fitting distance
to M87 (Ciardullo \etal 1998), convolved with the photometric error function,
and normalized to go through the point at $m_{5007} = 26.9$.  The excess of
$\sim 10$ planetaries at the bright end of the luminosity function 
demonstrates the presence of intracluster stars.}
\end{figure}

\clearpage
\begin{deluxetable}{lrrcccc}
\tablenum{1}
\tablewidth{0pt}
\tablecaption{Record of Observations }
\tablehead{
   &   &   &\colhead{Total Exposure} &\colhead{Mean} &\colhead{Mean} 
&\colhead{m$_{5007}$ completeness} \\
\colhead{Field }           & \colhead{$\alpha (2000)$}      &
\colhead{$\delta (2000)$} & \colhead{Time (s)} & 
\colhead{Transparency} & \colhead{Seeing } & 
\colhead{limit}} 

\startdata
1	& 12 27 47.51 & 12 42 24.29 & 10800 & 78\%  & $1\parcsec 1$ & 26.4 \nl
2	& 12 29 49.54 &  8 34 22.35 & 18000 & 100\% & $1\parcsec 5$ & 26.8 \nl
3	& 12 30 47.50 & 12 37 14.38 & 18000 & 98\%  & $1\parcsec 4$ & 27.0 \nl 
\enddata
\end{deluxetable}

\begin{deluxetable}{llccccc}
\tablenum{2}
\tablewidth{0pt}
\tablecaption{Model Results}
\tablehead{
\colhead{}  &\colhead{}  &\colhead{}  &\colhead{Implied} &\colhead{} 
&\colhead{} &\colhead{} \\
\colhead{}  &\colhead{}  &\colhead{$\alpha_{2.5}$}  &\colhead{Luminosity} 
&\colhead{}  &\colhead{L$_{\it galaxies}$} &\colhead{} \\
\colhead{Field}  &\colhead{Model} &\colhead{[$10^{-9}$~PN-$L_{\odot}^{-1}$]} 
&\colhead{[$\times 10^{9} L\subsun$ per field]} 
&\colhead{$\mu_{B}$} &\colhead{[$\times 10^{9} L\subsun$ per field]} 
& \colhead{Fraction}} 

\startdata
1 & symmetric cluster & 20 & 55  & 25.5 & 7.1  & 88\%\nl
1 & single component  & 20 & 20  & 26.6 & 7.1  & 74\%\nl
1 & single component  & 50 & 5   & 28.1 & 7.1 & 41\% \nl
2 & symmetric cluster & 20 & 15  & 27.0 & $\leq$ 7.1 & $ \geq 68\% $ \nl      
2 & single component & 20  & 5   & 28.1 & $\leq$ 7.1 & $ \geq 41\% $ \nl
2 & single component & 50  & 1   & 29.9 & $\leq$ 7.1 & $ \geq 12\% $ \nl 
3 & single component & 20  & 2   & 29.1 & 5.8   & $ 26\% $ \nl 
3 & single component & 50  & 0.8 & 30.1 & 5.8   & 12\%\nl
\enddata
\end{deluxetable}

\end{document}